# Energy-Efficient Management of Data Center Resources for Cloud Computing: A Vision, Architectural Elements, and Open Challenges

Rajkumar Buyya[1,2], Anton Beloglazov[1], and Jemal Abawajy[3]

[1]**Clou**d Computing and **D**istributed **S**ystems (CLOUDS) Laboratory
Department of Computer Science and Software Engineering
The University of Melbourne, Australia

[2] Manjrasoft Pty Ltd, Australia

[3]School of Information Technology
Deakin University, Melbourne, Australia

{raj, abe}@csse.unimelb.edu.au, jemal.abawajy@deakin.edu.au

**Abstract**

Cloud computing is offering utility-oriented IT services to users worldwide. Based on a pay-as-you-go model, it enables hosting of pervasive applications from consumer, scientific, and business domains. However, data centers hosting Cloud applications consume huge amounts of energy, contributing to high operational costs and carbon footprints to the environment. Therefore, we need Green Cloud computing solutions that can not only save energy for the environment but also reduce operational costs. This paper presents vision, challenges, and architectural elements for energy-efficient management of Cloud computing environments. We focus on the development of dynamic resource provisioning and allocation algorithms that consider the synergy between various data center infrastructures (i.e., the hardware, power units, cooling and software), and holistically work to boost data center energy efficiency and performance. In particular, this paper proposes (a) architectural principles for energy-efficient management of Clouds; (b) energy-efficient resource allocation policies and scheduling algorithms considering quality-of-service expectations, and devices power usage characteristics; and (c) a novel software technology for energy-efficient management of Clouds. We have validated our approach by conducting a set of rigorous performance evaluation study using the CloudSim toolkit. The results demonstrate that Cloud computing model has immense potential as it offers significant performance gains as regards to response time and cost saving under dynamic workload scenarios.

## 1. Introduction
### 1.1. Computing Utilities, Data Centers and Cloud Computing: Vision and Potential

In 1969, Leonard Kleinrock [1], one of the chief scientists of the original Advanced Research Projects Agency Network (ARPANET) which seeded the Internet, said: *"As of now, computer networks are still in their infancy, but as they grow up and become sophisticated, we will probably see the spread of 'computer utilities' which, like present electric and telephone utilities, will service individual homes and offices across the country."* This vision of computing utilities based on a service provisioning model anticipated the massive transformation of the entire computing industry in the 21st century whereby computing services will be readily available on demand, like other utility services available in today's society. Similarly, users (consumers) need to pay providers only when they access the computing services. In addition, consumers no longer need to invest heavily or encounter difficulties in building and maintaining complex IT infrastructure.

In such a model, users access services based on their requirements without regard to where the services are hosted. This model has been referred to as *utility computing*, or recently as *Cloud computing* [5]. The latter term denotes the infrastructure as a "Cloud" from which businesses and users can access applications as services from anywhere in the world on demand. Hence, Cloud computing can be classified as a new paradigm for the dynamic provisioning of computing services supported by state-of-the-art data centers that usually employ Virtual Machine (VM) technologies for consolidation and environment isolation purposes [11]. Many computing service providers including Google, Microsoft, Yahoo, and IBM are rapidly deploying data centers in various locations around the world to deliver Cloud computing services. The potential of this trend can be noted from the statement:

- *"The Data Center Is The Computer,"* by Professor David Patterson of the University of California, Berkeley, an ACM Fellow, and former President of the ACM – CACM [2].

Cloud computing delivers infrastructure, platform, and software (applications) as services, which are made available to consumers as subscription-based services under the pay-as-you-go model. In industry these services are referred to as Infrastructure as a Service (IaaS), Platform as a Service (PaaS), and Software as a Service (SaaS) respectively. A recent Berkeley report [23] stated "Cloud Computing, the long-held dream of computing as a utility, has the potential to transform a large part of the IT industry, making software even more attractive as a service".

Clouds aim to drive the design of the next generation data centers by architecting them as networks of virtual services (hardware, database, user-interface, application logic) so that users can access and deploy applications from anywhere in the world on demand at competitive costs depending on their QoS (Quality of Service) requirements [3]. Developers with innovative ideas for new Internet services no longer require large capital outlays in hardware to deploy their service or human expense to operate it [23]. Cloud computing offers significant benefits to IT companies by freeing them from the low-level task of setting up basic hardware and software infrastructures and thus enabling focus on innovation and creating business value for their services.

The business potential of Cloud computing is recognised by several market research firms. According to Gartner, Cloud market opportunities in 2013 will be worth $150 billion. Furthermore, many applications making use of utility-oriented computing systems such as Clouds emerge simply as catalysts or market makers that bring buyers and sellers together. This creates several trillion dollars worth of business opportunities to the utility/pervasive computing industry as noted by Sun co-founder Bill Joy [24]. He said "It would take time until these markets mature to generate this kind of value. Predicting now which companies will capture the value is impossible. Many of them have not even been created yet."

## 1.2 Cloud Infrastructure: Challenges and Requirements

Modern data centers, operating under the Cloud computing model are hosting a variety of applications ranging from those that run for a few seconds (e.g. serving requests of web applications such as e-commerce and social networks portals with transient workloads) to those that run for longer periods of time (e.g. simulations or large data set processing) on shared hardware platforms. The need to manage multiple applications in a data center creates the challenge of on-demand resource provisioning and allocation in response to time-varying workloads. Normally, data center resources are statically allocated to applications, based on peak load characteristics, in order to maintain isolation and provide performance guarantees. Until recently, high performance has been the sole concern in data center deployments and this demand has been fulfilled without paying much attention to energy consumption. The average data center consumes as much energy as 25,000 households [20]. As energy costs are increasing while availability dwindles, there is a need to shift focus from optimising data center resource management for pure performance to optimising for energy efficiency while maintaining high service level performance.

- *"The total estimated energy bill for data centers in 2010 is $11.5 billion and energy costs in a typical data center double every five years"*, according to McKinsey report [19].

Data centers are not only expensive to maintain, but also unfriendly to the environment. Data centers now drive more in carbon emissions than both Argentina and the Netherlands [20]. High energy costs and huge carbon footprints are incurred due to massive amounts of electricity needed to power and cool numerous servers hosted in these data centers. Cloud service providers need to adopt measures to ensure that their profit margin is not dramatically reduced due to high energy costs. For instance, Google, Microsoft, and Yahoo are building large data centers in barren desert land surrounding the Columbia River, USA to exploit cheap and reliable hydroelectric power [4]. There is also increasing pressure from Governments worldwide to reduce carbon footprints, which have a significant impact on climate change. For example, the Japanese government has established the Japan Data Center Council to address the soaring energy consumption of data centers [6]. Leading computing service providers have also recently formed a global consortium known as The Green Grid [7] to promote energy efficiency for data centers and minimise their environmental impact.

Lowering the energy usage of data centers is a challenging and complex issue because computing applications and data are growing so quickly that increasingly larger servers and disks are needed to process them fast enough within the required time period. **Green Cloud computing** is envisioned to achieve not only efficient processing and utilisation of computing infrastructure, but also minimise energy consumption. This is essential for ensuring that the future growth of Cloud computing is sustainable. Otherwise, Cloud computing with increasingly pervasive front-end client devices interacting with back-end data centers will cause an enormous escalation of energy usage. To address this problem, data center resources need to be managed in an energy-efficient manner to drive Green Cloud computing. In particular, Cloud resources need to be allocated not only to satisfy QoS requirements specified by users via Service Level Agreements (SLA), but also to reduce energy usage.

### 1.3 Our Contributions

Cloud providers have to ensure that they can be flexible in their service delivery to meet various consumer requirements for their services, while keeping the consumers isolated from the underlying infrastructure. However, to support Green Cloud computing, providers also need to minimise the energy consumption of Cloud infrastructure, while enforcing service delivery. Rising energy cost is a highly potential threat as it increases the Total Cost of Ownership (TCO) and reduces the Return on Investment (ROI) of Cloud infrastructure setup by providers. However, the current state-of-the-art in Cloud infrastructure has no or limited consideration for supporting energy-aware service allocation that meets QoS needs of consumers and minimises energy costs to maximise ROI.

The main objective of this work is to initiate research and development of energy-aware resource allocation mechanisms and policies for data centers so that Cloud computing can be a more sustainable eco-friendly mainstream technology to drive commercial, scientific, and technological advancement for future generations. Specifically, our work aims to:

- define an architectural framework and principles for energy-efficient Cloud computing;
- investigate energy-aware resource provisioning and allocation algorithms that provision data center resources to client applications in a way that improves the energy efficiency of the data center, without violating the negotiated Service Level Agreements (SLA);

a) develop autonomic and energy-aware mechanisms that self-manage changes in the state of resources effectively and efficiently to satisfy service obligations and achieve energy efficiency;

b) investigate heterogeneous workloads of various types of Cloud applications and develop algorithms for energy-efficient mixing and mapping of VMs to suitable Cloud resources in addition to dynamic consolidation of VM resource partitions; and

c) implement a prototype system – incorporating the above mechanisms, and techniques – and deploy it within the state-of-the-art operational Cloud infrastructures with real world demonstrator applications.

## 2. Related Work

Current state-of-the-art Cloud infrastructure such as Amazon EC2 [12] neither support energy-efficient resource allocation that considers consumer preference for energy saving schemes, nor utilise sophisticated economic models to set the right incentives for consumers to reveal information about their service demand accurately. Hence, providers are not able to foster essential information exchange with consumers and therefore cannot attain efficient service allocation, which meets consumer needs and expectations with regards to their energy saving preference for Green Cloud computing.

Market-based resource management [13] has been proposed by researchers to manage allocations of computing resources since it is effectively utilised in the field of economics to regulate supply and demand of limited goods. With Cloud computing emphasizing on a *pay-per-use* economic model, there is a high potential to apply market-based resource management techniques that justify the monetary return and opportunity cost of resource allocation according to consumer QoS expectations and baseline energy costs.

There are many proposed systems utilising market-based resource management for various computing areas [8], but none of these systems focus on the problem of energy efficiency in addition to maximizing profit. Thus, these systems are not able to support Green Cloud computing. They do not offer economic incentives that can encourage, discourage, or vary quality expectations of service requests from consumers with respect to energy saving schemes. A Green Cloud computing model not only facilitates providers to allocate service allocations efficiently to meet the customized needs of consumers, but also maximises revenue intake through more precise and differentiated pricing and utility policies based on specific consumer profiles. The potential of such a model was demonstrated in our own work in 2007 [9] and others in 2009 [10] in the context of HPC systems.

Existing energy-efficient resource allocation solutions proposed for various computing systems [14][15][16] cannot be implemented for Green Cloud computing. This is because they only focus on minimising energy consumption or their costs, and do not consider dynamic service requirements of consumers that can be changed on demand in Cloud computing environments. Hence, they do not emphasize autonomic energy-aware resource management mechanisms and policies exploiting VM resource allocation which is the main operating technology in Cloud Computing.

## 3. Green Cloud Architectural Elements

The aim of this paper is to addresses the problem of enabling energy-efficient resource allocation, hence leading to Green Cloud computing data centers, to satisfy competing applications' demand for computing services and save energy. Figure 1 shows the high-level architecture for supporting energy-efficient service allocation in Green Cloud computing infrastructure. There are basically four main entities involved:

a) *Consumers/Brokers***:** Cloud consumers or their brokers submit service requests from anywhere in the world to the Cloud. It is important to notice that there can be a difference between Cloud consumers and users of deployed services. For instance, a consumer can be a company deploying a Web application, which presents varying workload according to the number of "users" accessing it.

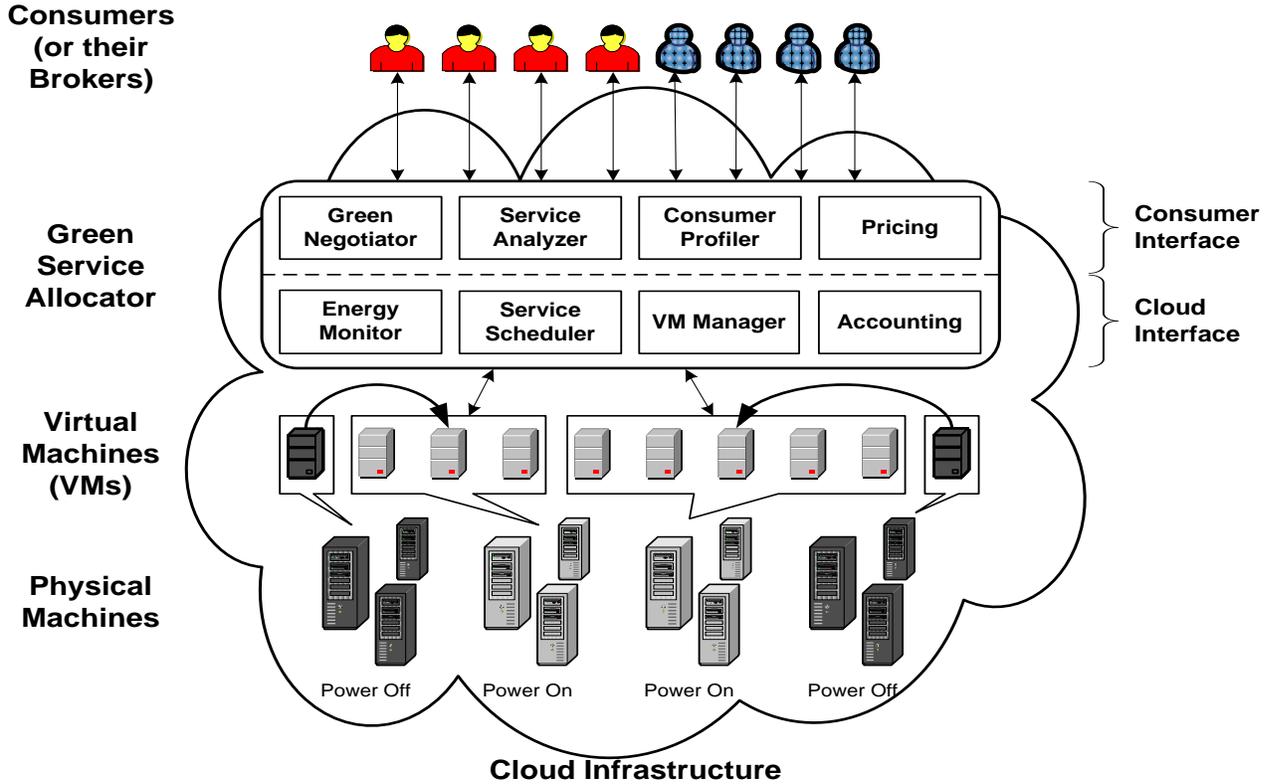

**Figure 1:** High-level system architectural framework.

b) *Green Resource Allocator***:** Acts as the interface between the Cloud infrastructure and consumers. It requires the interaction of the following components to support energy-efficient resource management:

- *Green Negotiator*: Negotiates with the consumers/brokers to finalize the SLA with specified prices and penalties (for violations of SLA) between the Cloud provider and consumer depending on the consumer's QoS requirements and energy saving schemes. In case of Web applications, for instance, QoS metric can be 95% of requests being served in less than 3 seconds.

- *Service Analyser*: Interprets and analyses the service requirements of a submitted request before deciding whether to accept or reject it. Hence, it needs the latest load and energy information from VM Manager and Energy Monitor respectively.

- *Consumer Profiler*: Gathers specific characteristics of consumers so that important consumers can be granted special privileges and prioritised over other consumers.

- *Pricing*: Decides how service requests are charged to manage the supply and demand of computing resources and facilitate in prioritising service allocations effectively.

- *Energy Monitor*: Observes and determines which physical machines to power on/off.

- *Service Scheduler*: Assigns requests to VMs and determines resource entitlements for allocated VMs. It also decides when VMs are to be added or removed to meet demand.

- *VM Manager*: Keeps track of the availability of VMs and their resource entitlements. It is also in charge of migrating VMs across physical machines.
- *Accounting*: Maintains the actual usage of resources by requests to compute usage costs. Historical usage information can also be used to improve service allocation decisions.

c) **VMs:** Multiple VMs can be dynamically started and stopped on a single physical machine to meet accepted requests, hence providing maximum flexibility to configure various partitions of resources on the same physical machine to different specific requirements of service requests. Multiple VMs can also concurrently run applications based on different operating system environments on a single physical machine. In addition, by dynamically migrating VMs across physical machines, workloads can be consolidated and unused resources can be put on a low-power state, turned off or configured to operate at low-performance levels (e.g., using DVFS) in order to save energy.

d) **Physical Machines:** The underlying physical computing servers provide hardware infrastructure for creating virtualised resources to meet service demands.

### 3.5 Working Prototype

As discussed earlier, the project leads to the development of a software platform that supports energy-efficient management and allocation of Cloud data center resources. In order to reduce the cost of software engineering, we will extensively reuse existing Cloud middleware and associated technologies. We will leverage third party Cloud technologies and services offerings including (a) VM technologies such as open source Xen and commercial one from VMWare and (b) Amazon's Elastic Computing Cloud (EC2), Simple Storage Service (S3); Microsoft's Azure. We will also leverage our own technologies such as Aneka, which is .NET-based platform for building enterprise Clouds. We will implement a generic resource manager and plug-in software adaptors to allow interaction with different Cloud management systems such as Aneka and Amazon EC2 engine. We will implement an energy-aware VM manager that guides the reallocation of VMs according to current resource requirements and states of physical nodes.

### 4. Energy-Aware Data Centre Resource Allocation

The problem of VM allocation can be divided in two: the first part is admission of new requests for VM provisioning and placing the VMs on hosts, whereas the second part is optimization of current allocation of VMs. The first part can be seen as a bin packing problem with variable bin sizes and prices. To solve it we apply modification of the Best Fit Decreasing (BFD) algorithm that is shown to use no more than 11/9 OPT + 1 bins (where OPT is the number of bins given by the optimal solution) [32]. In our modification (MBFD) we sort all VMs in decreasing order of current utilization and allocate each VM to a host that provides the least increase of power consumption due to this allocation. This allows leveraging heterogeneity of the nodes by choosing the most power-efficient ones. The complexity of the allocation part of the algorithm is $n \cdot m$, where $n$ is the number of VMs that have to be allocated and $m$ is the number of hosts.

Optimization of current allocation of VMs is carried out in two steps: at the first step we select VMs that need to be migrated, at the second step chosen VMs are placed on hosts using MBFD algorithm. We propose four heuristics for choosing VMs to migrate. The first heuristic, Single Threshold (ST), is based on the idea of setting upper utilization threshold for hosts and placing VMs while keeping the total utilization of CPU below this threshold. The aim is to preserve free resources to prevent SLA violation due to consolidation in cases when utilization by VMs increases. At each time frame all VMs are reallocated using MBFD algorithm with additional condition of keeping the upper utilization threshold not violated. The new placement is achieved by live migration of VMs [33].

The other three heuristics are based on the idea of setting upper and lower utilization thresholds for hosts and keeping total utilization of CPU by all VMs between these thresholds. If the utilization of CPU for a host goes below the lower threshold, all VMs have to be migrated from this host and the host has to be switched off in order to eliminate the idle power consumption. If the utilization goes over the upper threshold, some VMs have to be migrated from the host to reduce utilization in order to prevent potential SLA violation. We propose three policies for choosing VMs that have to be migrated from the host.

- Minimization of Migrations (MM) – migrating the least number of VMs to minimise migration overhead.

- Highest Potential Growth (HPG) – migrating VMs that have the lowest usage of CPU relatively to the requested in order to minimise total potential increase of the utilization and SLA violation.

- Random Choice (RC) – choosing the necessary number of VMs by picking them according to a uniformly distributed random variable.

The complexity of the MM algorithm is proportional to the product of the number of over- and under-utilized hosts and the number of VMs allocated to these hosts. The results of a simulation-based evaluation of the proposed algorithms in terms of power consumption, SLA violations and the number of VM migrations are presented in Section 6.

## 5. Early Experiments and Results

In this section, we will discuss some of our early performance analysis of the energy-aware allocation heuristics described in the previous section. As the targeted system is a generic Cloud computing environment, it is essential to evaluate it on a large-scale virtualised data center infrastructure. However, it is difficult to conduct large-scale experiments on a real infrastructure, especially when it is necessary to repeat the experiment with the same conditions (e.g. when comparing different algorithms). Therefore, simulations have been chosen as a way to evaluate the proposed heuristics. The CloudSim toolkit [34] has been chosen as a simulation platform as it is a modern simulation framework aimed at Cloud computing environments. In contrast to alternative simulation toolkits (e.g. SimGrid, GandSim), it supports modeling of on-demand virtualization enabled resource and application management. It has been extended in order to enable power-aware simulations as the core framework does not provide this capability. Apart from the power consumption modeling and accounting, the ability to simulate service applications with variable over time workload has been incorporated.

There are a few assumptions that have been made to simplify the model of the system and enable simulation-driven evaluation. The first assumption is that the overhead of VM migration is considered as negligible. Modeling the cost of migration of VMs is another research problem and is being currently investigated [33]. However, it has been shown that application of live migration of VMs can provide reasonable performance overhead. Moreover, with advancements of virtualization technologies, the efficiency of VM migration is going to be improved. Another assumption is that due to unknown types of applications running on VMs, it is not possible to build the exact model of such a mixed workload [31]. Therefore, rather than simulating particular applications, the utilization of CPU by a VM is generated as a uniformly distributed random variable. In the simulations we have defined that SLA violation occurs when a VM cannot get amount of MIPS that are requested. This can happen in cases when VMs sharing the same host require higher CPU performance that cannot be provided due to consolidation. To compare efficiency of the algorithms we use a characteristic called SLA violation percentage, or simply SLA violation, which is defined as a percentage of SLA violation events relatively to the total number of measurements.

### 5.1 Power Model

Power consumption by computing nodes in data centers consists of consumption by CPU, disk storage and network interfaces. In comparison to other system resources, CPU consumes larger amount of energy, and hence in this work we focus on managing its power consumption and efficient usage.

Recent studies [28], [29], [30], [31] show that application of DVFS on CPU results in almost linear power-to-frequency relationship. The reason lies in the limited number of states that can be set to the frequency and voltage of CPU and the fact that DVFS is not applied to other system components apart from CPU. Moreover, these studies show that in average an idle server consumes approximately 70% of the power consumed by the server running at full CPU speed. This fact justifies the technique of switching idle servers off to reduce total power consumption. Therefore, in this work we use power model defined in (1).

$$P(u) = k * P_{max} + (1-k) * P_{max} * u. \qquad (1)$$

where $P_{max}$ is the maximum power consumed when the server is fully utilised; $k$ is the fraction of power consumed by the idle server; and $u$ is the CPU utilisation. The utilisation of CPU may change over time due to variability of the workload. Thus, the CPU utilization is a function of time and represented as $u(t)$. Therefore, total energy (E) consumption by a physical node can be defined as an integral of the power consumption function over a period of time (2).

$$E = \int_t P(u(t)). \qquad (2)$$

### 5.2 Experimental Setup

We simulated a data center that comprises 100 heterogeneous physical nodes. Each node is modeled to have one CPU core with performance equivalent to 1000, 2000 or 3000 Million Instructions Per Second (MIPS), 8 Gb of RAM and 1 TB of storage. Power consumption by the hosts is defined according to the model described in Section 5.1. According to this model, a host consumes from 175 W with 0% CPU utilization and up to 250 W with 100% CPU utilization. Each VM requires one CPU core with 250, 500, 750 or 1000 MIPS, 128 MB of RAM and 1 GB of storage. The users submit requests for provisioning of 290 heterogeneous VMs that fills the full capacity of the simulated data center. Each VM runs a web-application or any kind of application with variable workload, which is modeled to create the utilization of CPU according to a uniformly distributed random variable. The application runs for 150,000 MIPS that equals to 10 minutes of execution on 250 MIPS CPU with 100% utilization. Initially, VMs are allocated according to the requested characteristics assuming 100% utilization. Each experiment has been run 10 times and the presented results are built upon the mean values.

## 5.3 Simulation Results

For the benchmark experimental results we have used a Non Power Aware (NPA) policy. This policy does not apply any power aware optimizations and implies that all hosts run at 100% CPU utilization and consume maximum power. The second policy applies DVFS, but does not perform any adaptation of allocation of VMs in run-time. For the simulation setup described above, using the NPA policy leads to the total energy consumption of 9.15 KWh, whereas DVFS allows decreasing this value to 4.4 KWh.

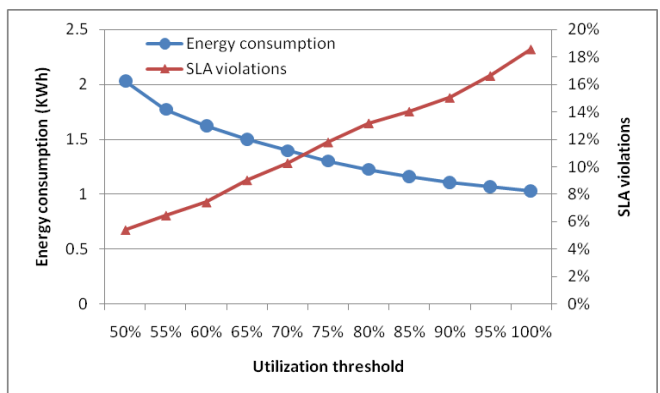

**Figure 2.** Energy consumption and SLA violations by ST policy.

To evaluate ST policy we conducted several experiments with different values of the utilization threshold. The simulation results are presented in Figure 2. The results show that energy consumption can be significantly reduced relatively to NPA and DVFS policies – by 77% and 53% respectively with 5.4% of SLA violations. They show that with the growth of the utilization threshold energy consumption decreases, whereas percentage of SLA violations increases. This is due to the fact that higher utilization threshold allows more aggressive consolidation of VMs, however, by the cost of the increased risk of SLA violations.

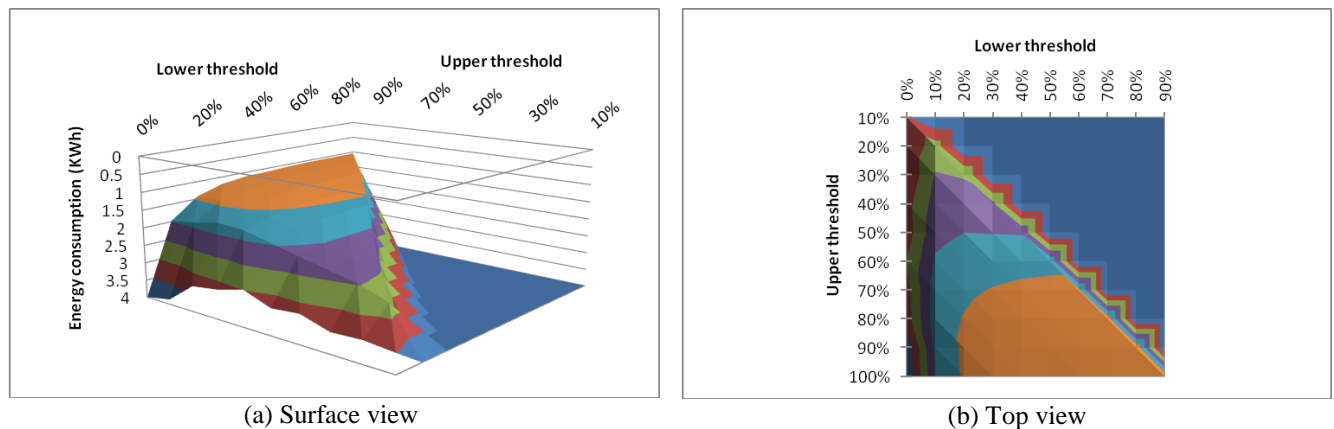

(a) Surface view        (b) Top view

**Figure 3.** Energy to thresholds relationship for MM policy

To evaluate two-threshold policies it is necessary to determine the best values for the utilization thresholds in terms of power consumption and QoS provided. Therefore, at first we simulated MM policy with different values of thresholds varying absolute values of the thresholds as well as the interval between the lower and upper thresholds. The results showing the energy consumption achieved by using this policy are presented in Figure 3. The lowest values of energy consumption can be gained by setting the lower threshold from 10% to 90% and the upper threshold from 50% to 100%. However, the obtained intervals of the thresholds are wide. Therefore, to determine the concrete values, we have compared the thresholds by the percentage of SLA violations caused, as rare SLA violations ensure high QoS. The experimental results have shown that the minimum values of both characteristics can be achieved using 40% as the interval between the utilization thresholds.

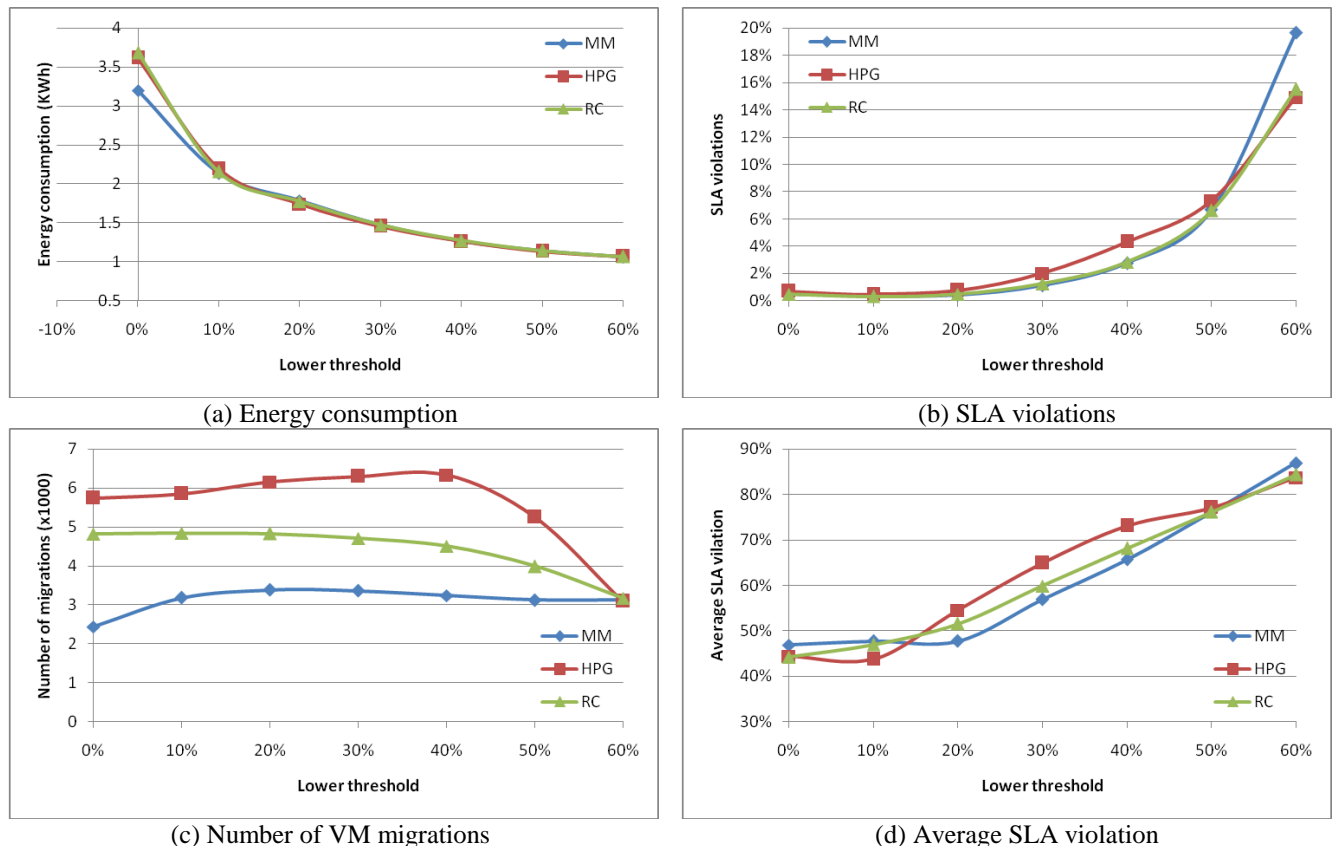

(a) Energy consumption  (b) SLA violations

(c) Number of VM migrations  (d) Average SLA violation

**Figure 4.** Comparison of two-threshold algorithms.

We have compared MM policy with HPG and RC policies varying exact values of the thresholds but preserving 40% interval between them. The results (Figure 4) show that these policies allow the achievement of approximately the same values of energy consumption and SLA violations. Whereas the number of VM migrations produced by MM policy is reduced in comparison to HPG policy by maximum of 57% and 40% on average and in comparison to RC policy by maximum of 49% and 27% on average.

**Table 1.** The final simulation results.

| Policy | Energy, kWh | SLA, % | VM migrations | Avg. SLA, % |
|---|---|---|---|---|
| NPA | 9.15 | - | - | - |
| DVFS | 4.40 | - | - | - |
| ST 50% | 2.03 | 5.41 | 35 226 | 81 |
| ST 60% | 1.50 | 9.00 | 34 231 | 89 |
| MM 30-70% | 1.48 | 1.11 | 3 359 | 56 |
| MM 40-80% | 1.27 | 2.75 | 3 241 | 65 |
| MM 50-90% | 1.14 | 6.69 | 3 120 | 76 |

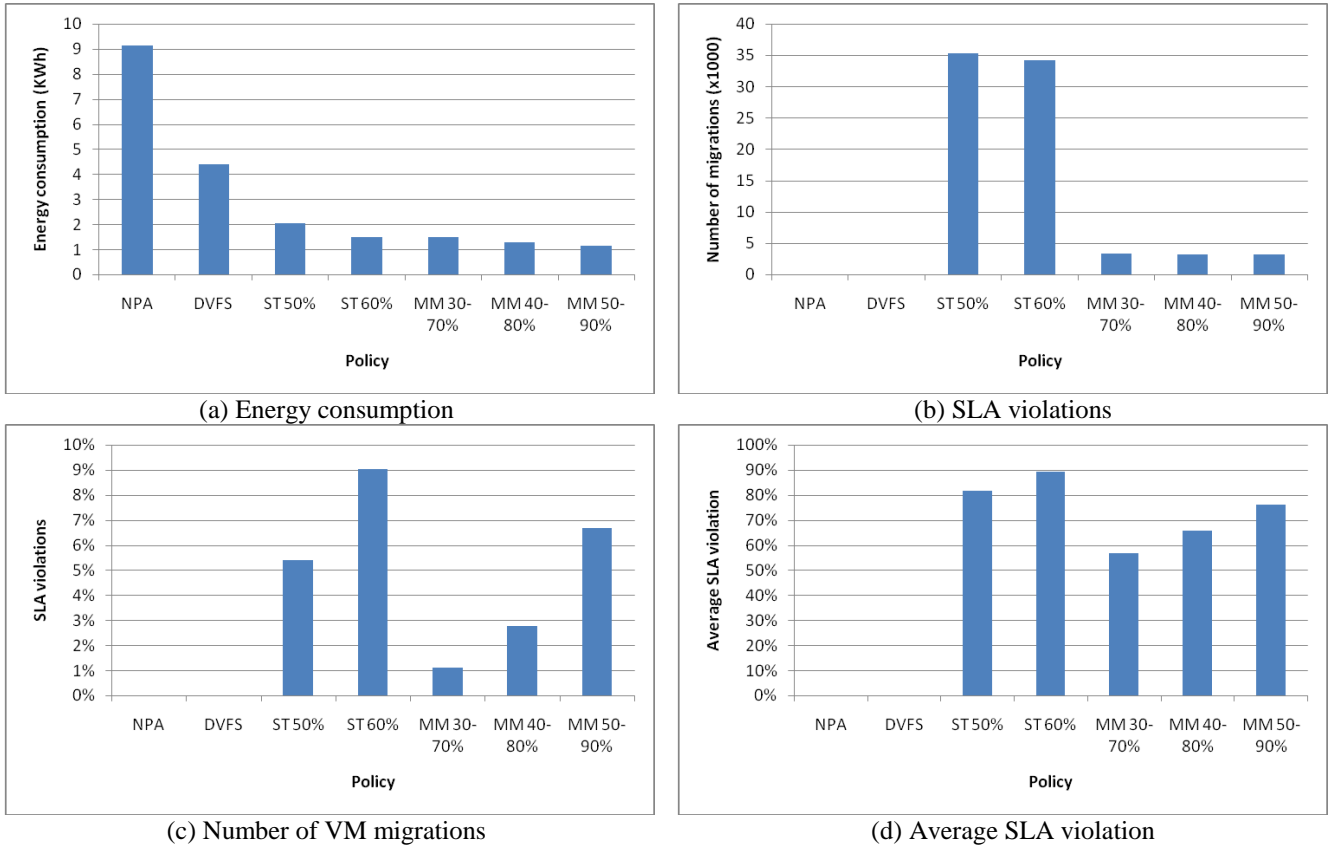

**Figure 5.** The final simulation results.

Final results comparing all the policies with different values of the thresholds are presented in Table 1 and in Figure 5. The results show that dynamic reallocation of VMs according to current utilization of CPU provides higher energy savings compared to static allocation policies. MM policy leads to the best energy savings: by 83%, 66% and 23% less energy consumption relatively to NPA, DVFS and ST policies respectively with thresholds 30-70% and ensuring percentage of SLA violations of 1.1%; and by 87%, 74% and 43% with thresholds 50-90% and 6.7% of SLA violations. MM policy leads to more than 10 times less VM migrations than ST policy. The results show flexibility of the algorithm, as the thresholds can be adjusted according to SLA requirements. Strict SLA (1.11%) allow the achievement of the energy consumption of 1.48 KWh. However, if SLA are relaxed (6.69%), the energy consumption is further reduced to 1.14 KWh.

## 6. Open Challenges

In this section, we identify key open problems that can be addressed at the level of management of system resources. Virtualisation technologies, which Cloud computing environments heavily rely on, provide the ability to transfer VMs between physical nodes using live of offline migration. This enables the technique of dynamic consolidation of VMs to a minimal number of nodes according to current resource requirements. As a result, the idle nodes can be switched off or put to a power saving mode (e.g. sleep, hibernate) to reduce total energy consumption by the data center. In order to validate the approach, we have proposed several resource allocation algorithms discussed in Section 4 and evaluated them by extensive simulation studies presented in Section 5. Despite the energy savings, aggressive consolidation of VMs may lead to a performance degradation and, thus result in SLA violation. Our resource management algorithms effectively address the trade-off between energy consumption and performance delivered by the system.

### 6.1 Energy-aware Dynamic Resource Allocation

Recent developments in virtualisation have resulted in its proliferation of usage across data centers. By supporting the movement of VMs between physical nodes, it enables dynamic migration of VMs according to QoS requirements. When VMs do not use all provided resources, they can be logically resized and consolidated on a minimal number of physical nodes, while idle nodes can be switched off.

Currently, resource allocation in a Cloud data center aims to provide high performance while meeting SLA, without a focus on allocating VMs to minimise energy consumption. To explore both performance and energy efficiency, three crucial issues must be addressed. First, excessive power cycling of a server could reduce its reliability. Second, turning resources off in a dynamic environment is risky from a QoS prospective. Due to the variability of the workload and aggressive consolidation, some VMs may not obtain required resources under peak load, so failing to meet the desired QoS. Third, ensuring SLA brings challenges to accurate application performance management in virtualized environments.

A virtual machine cannot exactly record the timing behaviour of a physical machine. This leads to the timekeeping problems resulting in inaccurate time measurements within the virtual machine, which can lead to incorrect enforcement of SLA. All these issues require effective consolidation policies that can minimise energy consumption without compromising the used-specified QoS requirements. To achieve this goal, we will develop novel QoS-based resources selection algorithms and mechanisms that optimise VM placements with the objective of minimizing communication overhead as described below.

## 6.2 QoS-based Resource Selection and Provisioning

Data center resources may deliver different levels of performance to their clients; hence, QoS-aware resource selection plays an important role in Cloud computing. Additionally, Cloud applications can present varying workloads. It is therefore essential to carry out a study of Cloud services and their workloads in order to identify common behaviors, patterns, and explore load forecasting approaches that can potentially lead to more efficient resource provisioning and consequent energy efficiency. In this context, we will research sample applications and correlations between workloads, and attempt to build performance models that can help explore the trade-offs between QoS and energy saving. Further, we will investigate a new online approach to the consolidation strategy of a data center that allows a reduction in the number of active nodes required to process a variable workload without degrading the offered service level. The online method will automatically select a VM configuration while minimising the number of physical hosts needed to support it. Moreover, another goal is to provide the broker (or consumers) with resource-selection and workload-consolidation policies that exploit the trade-offs between performance and energy saving.

## 6.3 Optimisation of Virtual Network Topologies

In virtualised data centers VMs often communicate between each other, establishing virtual network topologies. However, due to VM migrations or non-optimised allocation, the communicating VMs may end up hosted on logically distant physical nodes providing costly data transfer between each other. If the communicating VMs are allocated to the hosts in different racks or enclosures, the network communication may involve network switches that consume significant amount of power. To eliminate this data transfer overhead and minimise power consumption, it is necessary to observe the communication between VMs and place them on the same or closely located nodes. To provide effective reallocations, we will develop power consumption models of the network devices and estimate the cost of data transfer depending on the traffic volume. As migrations consume additional energy and they have a negative impact on the performance, before initiating the migration, the reallocation controller has to ensure that the cost of migration does not exceed the benefit.

## 6.4 Autonomic Optimisation of Thermal states and Cooling System Operation

A significant part of electrical energy consumed by computing resources is transformed into heat. High temperature leads to a number of problems, such as reduced system reliability and availability, as well as decreased lifetime of devices. In order to keep the system components within their safe operating temperature and prevent failures and crashes, the emitted heat must be dissipated. The cooling problem becomes extremely important for modern blade and 1-unit rack servers that lead to high density of computing resources and complicate heat dissipation. For example, in a 30,000 $ft^2$ data center with 1000 standard computing racks, each consuming 10kW, the initial cost of purchasing and installing the infrastructure is $2-$5 million; whereas the annual costs for cooling is around $4-$8 million [22]. Therefore, apart from the hardware improvements, it is essential to optimise the cooling system operation from the software side. There has been work on modelling thermal topology of a data center that can lead to more efficient workload placement [27]. The new challenges include how and when to reallocate VMs to minimise power drawn by the cooling system, while preserving safe temperature of the resources and minimising migration overhead and performance degradation.

We will investigate and develop a new thermal-management algorithm that monitors the thermal state of physical nodes and reallocate workload (VMs) from the overheated nodes to other nodes. In this case, the cooling systems of heated nodes can be slowed down, allowing natural power dissipation. We will develop an approach that leverages the temperature variations between different workloads, and swap them at an appropriate time to control the temperature. In addition, hardware level thermal management techniques, such as Dynamic Voltage and Frequency Scaling (DVFS) of modern processors, can lower the temperature when it surpasses the thermal threshold. As noted in our recent work these mechanisms can be effectively used whenever the QoS of hosted applications does not require processors to operate in full capacity [9]. We will extend it for a case where multiple diverse applications with different QoS requirements share the system simultaneously.

### 6.5 Efficient Consolidation of VMs for Managing Heterogeneous Workloads

Cloud infrastructure services provide users with the ability to provision virtual machines and allocate any kind of applications on them. This leads to the fact that different types of applications (e.g., enterprise, scientific, and social network applications) can be allocated on one physical computer node. However, it is not obvious how these applications can influence each other, as they can be data, network or compute intensive thus creating variable or static load on the resources. The problem is to determine what kind of applications can be allocated to a single host that will provide the most efficient overall usage of the resources. Current approaches to energy efficient consolidation of VMs in data centers do not investigate the problem of combining different types of workload. These approaches usually focus on one particular workload type or do not consider different kinds of applications assuming uniform workload. In contrast to the previous work, we propose an intelligent consolidation of VMs with different workload types. A compute intensive (scientific) application can be effectively combined with a web-application (file server), as the former mostly relies on CPU performance, whereas the latter utilises disk storage and network bandwidth. We will investigate which particular kind of applications can be effectively combined and what parameters influence the efficiency; and develop resource allocation algorithms for managing them. This knowledge can be applied to energy efficient resource management strategies in data centers to achieve more optimal allocation of resources and, therefore, improve utilisation of resources and reduce energy consumption. For the resource providers, optimal allocation of VMs will result in higher utilisation of resources and, therefore, reduced operational costs. End-users will benefit from decreased prices for the resource usage.

### 7. Concluding Remarks and Future Directions

This work advances Cloud computing field in two ways. First, it plays a significant role in the reduction of data center energy consumption costs and thus helps to develop a strong, competitive Cloud computing industry. This is especially important in the context of Australia as a recent Frost & Sullivan's report shows that Australia is emerging as one of the preferred data center hubs among the Asia Pacific countries [25]. Second, consumers are increasingly becoming conscious about the environment. In Australia, a recent study shows that data centers represent a large and rapidly growing energy consumption sector of the economy and is a significant source of $CO_2$ emissions [26]. Reducing greenhouse gas emissions is a key energy policy focus of many countries including Australia. Therefore, we expect researchers world-wide to put in a strong thrust on open challenges identified in this paper in order enhance energy-efficient management of Cloud computing environments.